\documentclass[article, twocolumn, 
   aps, pra,
  amsmath,amssymb,
  longbibliography,
  ]{revtex4-1}
\usepackage{graphicx,color}
\usepackage{amsmath}
\usepackage{natbib}
\usepackage{epsfig}
\begin{document}

\title{\color{blue} Transport properties of Lennard-Jones fluids: Freezing density scaling along isotherms.}

\author{S. A. Khrapak}\email{Sergey.Khrapak@gmx.de}
\affiliation{Joint Institute for High Temperatures, Russian Academy of Sciences, 125412 Moscow, Russia}
\author{A. G. Khrapak}
\affiliation{Joint Institute for High Temperatures, Russian Academy of Sciences, 125412 Moscow, Russia}

\begin{abstract}
It is demonstrated that properly reduced transport coefficients (self-diffusion, shear viscosity, and thermal conductivity) of Lennard-Jones fluids along isotherms exhibit quasi-universal scaling on the density divided by its value at the freezing point. Moreover, this scaling is closely related to the density scaling of transport coefficients of hard-sphere fluids. The Stokes-Einstein relation without the hydrodynamic diameter is valid in the dense fluid regime. The lower density boundary of its validity can serve as a practical demarcation line between gas-like and liquid-like regimes.
 
\end{abstract}

\date{\today}

\maketitle

\section{Introduction}

The Lennard-Jones (LJ) system is one of the most popular and extensively studied model systems in condensed matter physics. Many results on its transport properties have been published over the years. For recent reviews of simulation data available in the literature see e.g. Refs.~\cite{BellJPCB2019,HarrisJCP2020,AllersJCP2020}.   
Numerical data can be particularly useful in verifying and testing various models and approaches to the transport properties of liquids. The latter research topic remains important and timely despite considerable progress achieved over many decades~\cite{HansenBook,GrootBook,MarchBook}. 

Among the existing semi-quantitative models and scaling relationships perhaps the most familiar are the Stokes-Einstein relation between the self-diffusion and shear viscosity coefficients~\cite{FrenkelBook,ZwanzigJCP1983,Balucani1990,CostigliolaJCP2019,KhrapakMolPhys2019}, the excess entropy scaling of transport coefficients~\cite{RosenfeldPRA1977,DzugutovNature1996,
RosenfeldJPCM1999,DyreJCP2018}, density scaling~\cite{FragiadakisJCP2019,HarrisJCP2020} as well as the freezing temperature scaling~\cite{RosenfeldPRE2000,CostigliolaJCP2018,KhrapakAIPAdv2018}.   

Some of the scaling relationships can be rationalized within the framework of the isomorph theory~\cite{DyreJPCB2014,VeldhorstPoP2015,DyreJCP2018}. This theory predicts that many liquids exhibit an approximate
``hidden'' scale invariance that implies the existence of lines in the thermodynamic phase diagram, so-called isomorphs, along which structure and dynamics in properly reduced units
are invariant to a good approximation. Excess entropy is constant along isomorphs and this can serve as an explanation of excess energy scaling~\cite{DyreJCP2018}. Additionally, the melting and freezing lines are approximate isomorphs~\cite{PedersenNatCommun2016,SaijaJCP2001} and therefore the lines that are parallel to them in some vicinity should be also isomorphs. This explains the observation that the transport coefficients of some simple fluids behave quasi-universally when plotted versus temperature scaled by the freezing temperature. But this also suggests that the transport coefficients should coincide when plotted versus density scaled by the freezing density. The purpose of this paper is to prove this conjecture.

For some systems the freezing density scaling is trivial. A special case is the hard-sphere (HS) system. Since all the structural and dynamical properties depend on the packing fraction (reduced density) alone, the freezing density scaling is obvious. 
Another relevant example is related to strongly coupled Yukawa fluids,  which are often used to model classical systems of charged particles immersed in a neutralizing medium, such as colloidal suspensions and complex plasmas~\cite{FortovUFN,FortovPR,IvlevBook}. Transport coefficients of strongly coupled Yukawa fluids turn out to be some quasi-universal functions of the reduced coupling parameter $\Gamma/\Gamma_{\rm fr}$
~\cite{RosenfeldPRE2000,OhtaPoP2000,SaigoPoP2002,VaulinaPRE2002,VaulinaPoP2002,
FortovPRL2003,FaussurierPRE2003,DonkoPRE2004,VaulinaPRE2010,
KhrapakPoP2012,KhrapakJPCO2018}. Here the coupling parameter $\Gamma=Q^2\rho^{1/3}/T$ is the ratio of the Coulomb interaction energy at a mean interparticle separation ($Q$ being the particle charge and $\rho$ particle density) to the kinetic energy, characterized by the system temperature $T$ (measured in energy units) and $\Gamma_{\rm fr}$ is the value of $\Gamma$ at the freezing point (the coupling parameter at the melting point, $\Gamma_{\rm m}$, can be used as well because the fluid-solid coexistence region is very narrow~\cite{HynninenPRE2003} and one normally does not discriminate between freezing and melting points in Yukawa systems). The success of this scaling for Yukawa systems was put in the context of the isomorph theory~\cite{VeldhorstPoP2015}. For the present discussion it is important that a universal scaling with $\Gamma/\Gamma_{\rm fr}$ implies not only scaling with $T/T_{\rm fr}$, but also a scaling with $\rho/\rho_{\rm fr}$, so that the freezing density scaling emerges trivially.  


In this paper we demonstrate that the freezing density scaling is more general and is not limited to trivial examples given above. In particular, it works very well for sufficiently dense LJ fluids in a wide region of the LJ phase diagram. Simulation data on the transport coefficient of LJ fluids along isotherms fully comply with this scaling. Moreover, we demonstrate that the observed scaling is closely related to that of the hard-sphere fluid. It is shown that the Stokes-Einstein relation between the diffusion and viscosity coefficients is valid in the dense liquid regime, and that the boundary of this regime can be used to locate the crossover between gas-like and liquid-like regions on the phase diagram. 

\section{Formulation}

The LJ potential, which is often used to approximate interactions in liquefied noble gases, reads 
\begin{equation}
\phi(r)=4\epsilon\left[\left(\frac{\sigma}{r}\right)^{12}-\left(\frac{\sigma}{r}\right)^{6}\right], 
\end{equation}
where  $\epsilon$ and $\sigma$ are the energy and length scales (or LJ units), respectively. The density and  temperature in LJ units are $\rho_*=\rho\sigma^3$ and $T_*=T/\epsilon$. The LJ potential exhibits repulsion at short interatomic distances and attraction at long distances. As a result, the phase diagram of a LJ system contains a liquid-vapor phase transition with liquid-vapor coexistence region and a liquid-vapor critical point, in addition to the fluid-solid phase transition (systems of purely repulsive particles, such as for instance hard spheres, do not show a liquid phase).     

To demonstrate the validity and usefulness of the freezing density scaling we consider two sets of extensive numerical simulation data for the LJ transport coefficients. The viscosity and self-diffusion data have been published by Meier {\it et al.}~\cite{Meier2002,MeierJCP_1,MeierJCP_2}. The second data set for self-diffusion, viscosity, and thermal conductivity coefficients is due to Baidakov {\it et al.}~\cite{BaidakovFPE2011,BaidakovJCP2012,BaidakovJCP2014}.   
The data have been obtained in a wide region of the LJ phase diagram. Though simulations protocols were different, the two datasets are in good agreement where they overlap~\cite{HarrisJCP2020}. 

To be concrete, the viscosity and self-diffusion coefficients at $T_*=1, 1.5$ and 2 are taken from Refs.~\cite{BaidakovFPE2011,BaidakovJCP2012} and at $T_*=3$ and 4 from Ref.~\cite{Meier2002}. The thermal conductivity coefficients at $T_*=1, 1.5$ and 2 are taken from Ref.~\cite{BaidakovJCP2014}. We do not consider data points from the fluid-solid coexistence region.    

We reduce the transport coefficients (originally expressed in LJ units) using the so-called system-independent Rosenfeld normalization~\cite{RosenfeldJPCM1999}:
\begin{equation}\label{Rosenfeld}
D_{\rm R}  =  D\frac{\rho^{1/3}}{v_{\rm T}} , \quad
\eta_{\rm R}  =  \eta \frac{\rho^{-2/3}}{m v_{\rm T}}, \quad \lambda_{\rm R}  =  \lambda \frac{\rho^{-2/3}}{v_{\rm T}}, 
\end{equation}
where $\rho^{-1/3}=\Delta$ defines the characteristic interatomic separation, $v_{\rm T}=\sqrt{T/m}$ is the thermal velocity, $m$ is the atomic mass, and the temperature is expressed in energy units ($k_{\rm B}=1$).

In the first approximation, the transport coefficients of a dilute gas consisting of hard spheres of diameter $\sigma$ (HS gas) are~\cite{LifshitzKinetics}
\begin{eqnarray}\label{HSD}
D &=& \frac{3}{8\rho\sigma^2}\left(\frac{T}{\pi m}\right)^{1/2} \Rightarrow D_{\rm R}  =  \frac{3}{8\sqrt{\pi}\rho_*^{2/3}}\simeq\frac{0.212}{\rho_*^{2/3}} , \\ \label{HSeta}
\eta &=& \frac{5}{16\sigma^2}\left(\frac{mT}{\pi}\right)^{1/2} \Rightarrow \eta_{\rm R}  =  \frac{5}{16\sqrt{\pi}\rho_*^{2/3}}\simeq\frac{0.176}{\rho_*^{2/3}}, \\ \label{HSlambda}
\lambda &=& \frac{75}{64\sigma^2}\left(\frac{T}{\pi m}\right)^{1/2} \Rightarrow \lambda_{\rm R} = \frac{75}{64\sqrt{\pi}\rho_*^{2/3}}\simeq\frac{0.661}{\rho_*^{2/3}}.
\end{eqnarray}
Remarkably, when expressed in Rosenfeld's units, the transport coefficients of dilute hard-sphere gas exhibit universal density scaling: All transport coefficients decay as $\propto \rho_*^{-2/3}$. This density scaling is of course only approximate for dilute LJ gases. The actual transport cross sections are different from the hard-sphere model and specifics of scattering in the LJ potential has to be properly accounted for (see e.g. Refs.~\cite{HirschfelderBook,HirschfelderJCP1948,SmithJCP1964,
KhrapakPRE2014_Scattering,
KhrapakEPJD2014,Kristiansen2020} and references therein for some related works). 

Contrary to the situation with dilute gases no accurate general theory of transport process in liquids still exists, and that is why universal relationships and scalings are of considerable interest. The freezing density scaling along isotherms is illustrated in the next Section.

\section{Results and discussion}

\begin{figure}
\includegraphics[width=8.5cm]{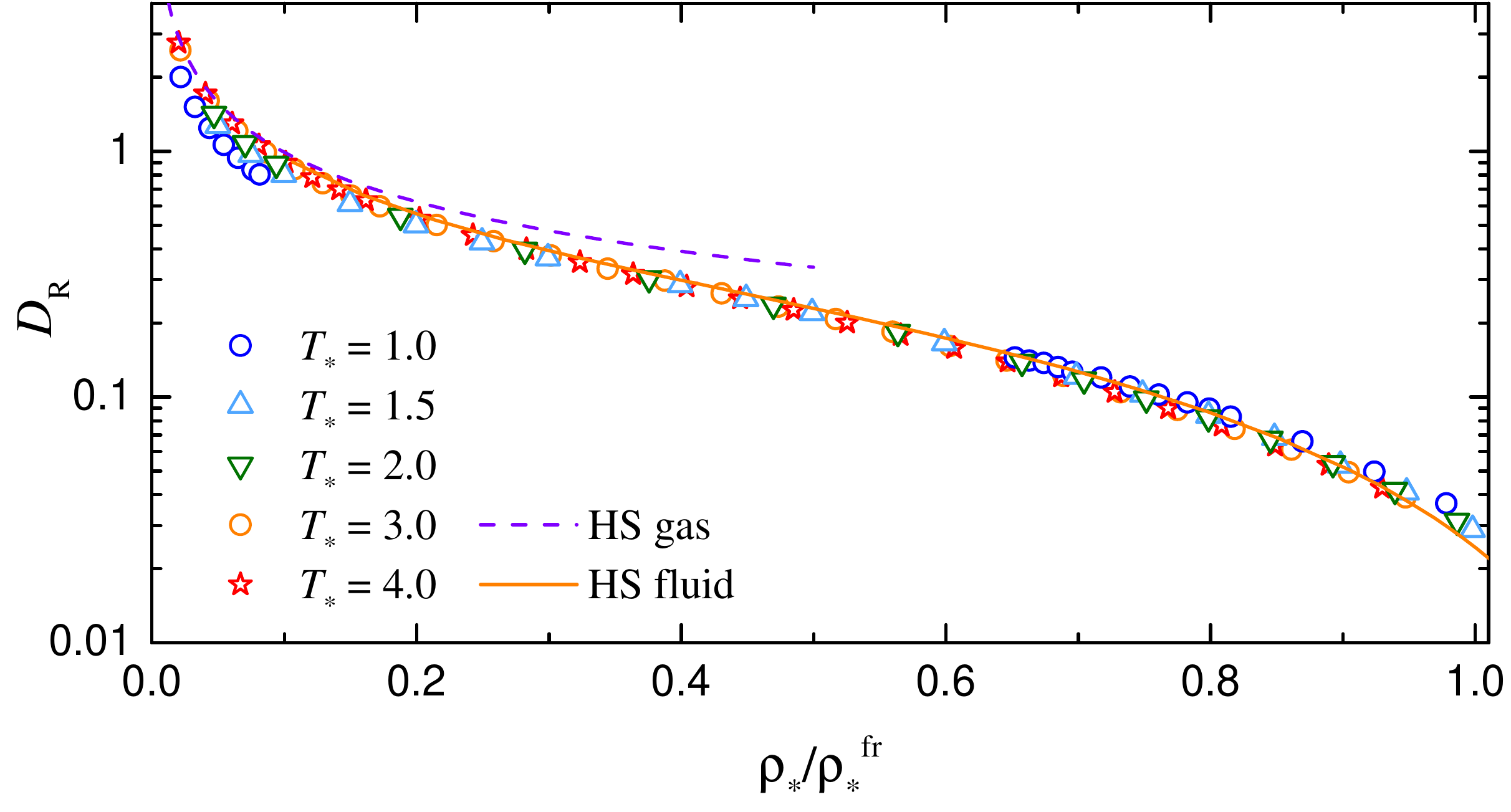}
\caption{(Color online) Reduced self-diffusion coefficient of a LJ fluid $D_{\rm R}$ vs the reduced density $\rho_*/\rho_*^{\rm fr}$. The symbols correspond to numerical data tabulated in Refs.~\cite{Meier2002,BaidakovFPE2011}. The dashed and solid curves denote the HS gas and fluid asymptotes. The latter is based on the simulation data from Ref.~\cite{Pieprzyk2019}.}
\label{Fig1}
\end{figure}

\begin{figure}
\includegraphics[width=8.5cm]{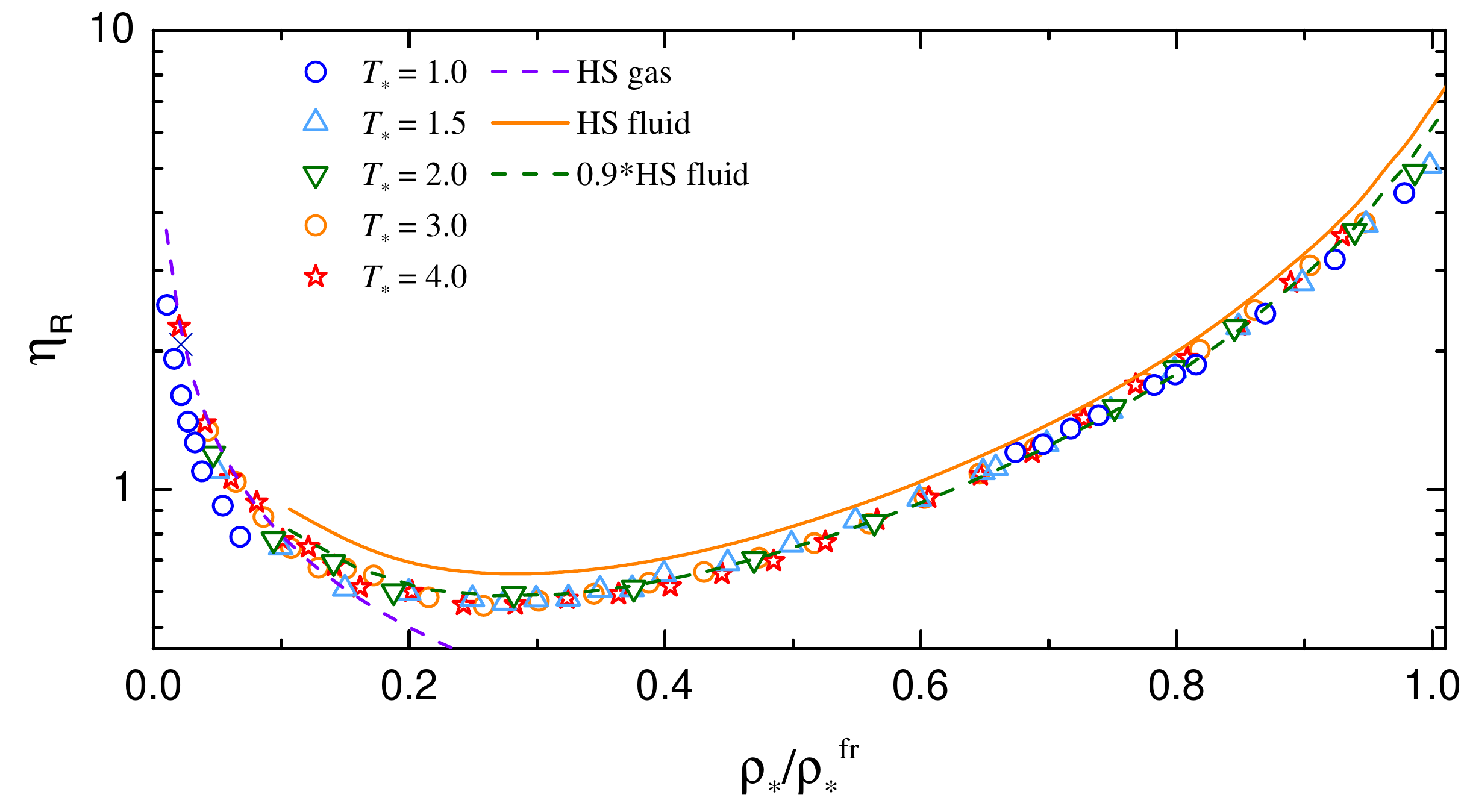}
\caption{(Color online) Reduced shear viscosity coefficient of a LJ fluid $\eta_{\rm R}$ vs the reduced density $\rho_*/\rho_*^{\rm fr}$. The symbols correspond to numerical data tabulated in Refs.~\cite{Meier2002,BaidakovJCP2012}. The dashed and solid curves denote the HS gas and fluid asymptotes. The latter is based on the simulation data from Ref.~\cite{Pieprzyk2019}.}
\label{Fig2}
\end{figure}

\begin{figure}
\includegraphics[width=8.5cm]{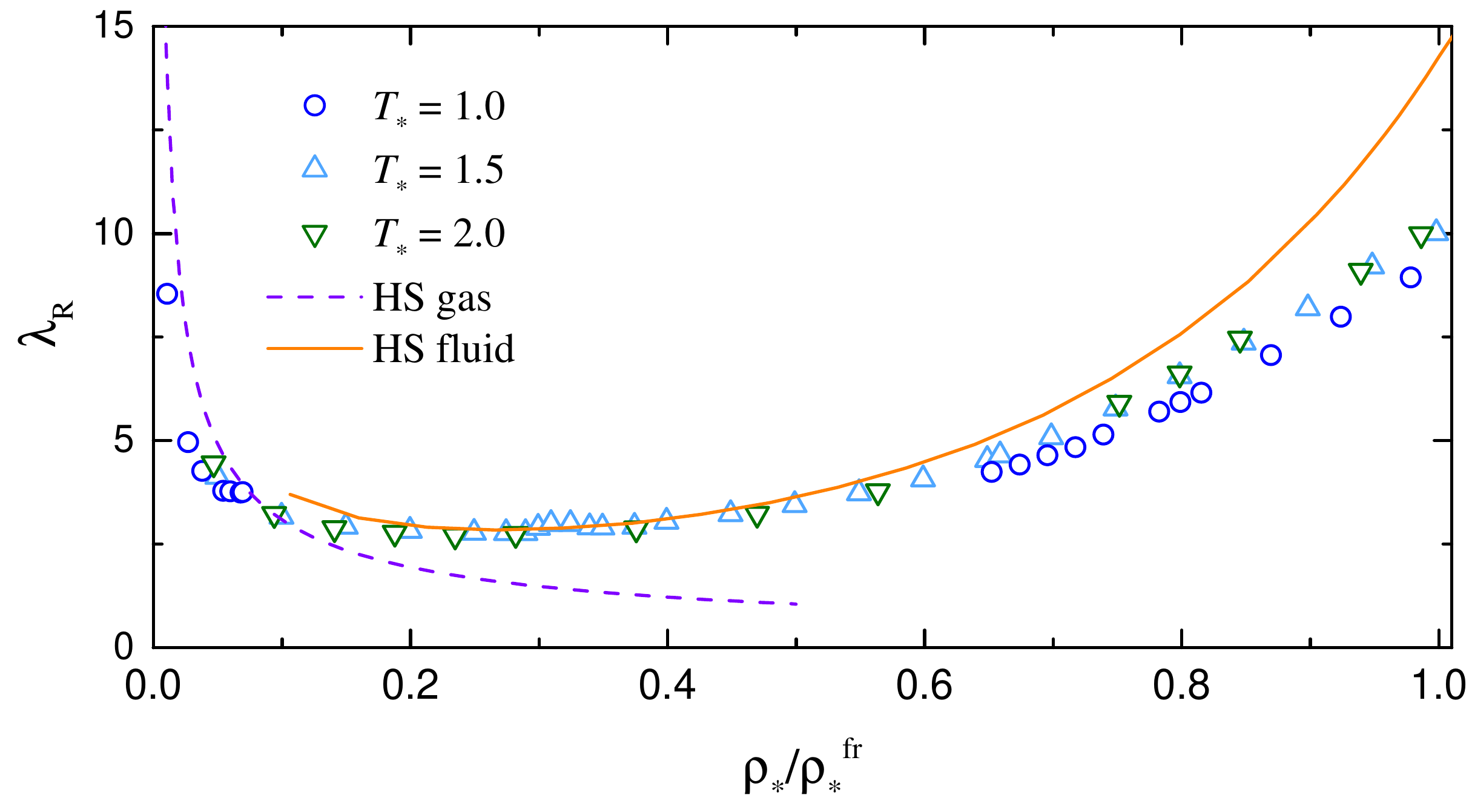}
\caption{(Color online) Reduced thermal conductivity coefficient of a LJ fluid $\lambda_{\rm R}$ vs the reduced density $\rho_*/\rho_*^{\rm fr}$. The symbols correspond to numerical data tabulated in Ref.~\cite{BaidakovJCP2014}. The dashed and solid curves denote the HS gas and fluid asymptotes. The latter is based on the simulation data from Ref.~\cite{Pieprzyk2020}.}
\label{Fig3}
\end{figure}

 Our main results are plotted in Figs.~\ref{Fig1} - \ref{Fig3}, where we plot the reduced transport coefficients versus the density normalized to its value at the freezing point, taken from Ref.~\cite{SousaJCP2012}. They demonstrate that the self-diffusion and viscosity data (Figs.~\ref{Fig1} and \ref{Fig2}) coincide excellently when plotted versus density reduced by its value at the freezing point. The coincidence is not so impressive for the thermal conductivity data shown in Fig.~\ref{Fig3}). It should be noted that the thermal conductivity coefficient is proportional to the specific heat $c_v$, which exhibits a critical enhancement near the critical point. Thus the behaviour of thermal conductivity on near-critical isotherms can be expected to be somewhat different from that on supercritical isotherms~\cite{KhrapakPRE2021}. Additionally, some scattering may  reflect the accuracy of numerical data. 

The quasi-universality of transport coefficients extend to very low density, down to $\rho_*/\rho_*^{\rm fr}\simeq 0.1$. For lower densities the freezing density scaling is not appropriate. Here the dilute gas regime is realised and the transport coefficients are determined by the corresponding transport cross sections~\cite{HirschfelderBook,KhrapakEPJD2014}.    

Two additional curves are plotted in Figs.~\ref{Fig1} - \ref{Fig3}. The dashed curves correspond to Eqs.~(\ref{HSD}) - (\ref{HSlambda}). As pointed out above, they do not represent accurate approximations for the LJ gas, but the main trends in the density dependence are very well reproduced. To plot these curves we assumed $\rho_{*}^{\rm fr}=1$, which is appropriate for the range of $T_*$ investigated due to relatively weak dependence of $\rho_*^{\rm fr}$ on $T_*$ (actually $\rho_{*}^{\rm fr}\simeq 1$ at $T_*\simeq 1.5$~\cite{SousaJCP2012}).
     
The solid curves correspond to the freezing density scaling of transport coefficients of a HS fluid. To plot these curves we have used extensive molecular simulation results by Pieprzyk {\it et al}.~\cite{Pieprzyk2019,Pieprzyk2020}. The use of large simulation systems and long simulation times allowed accurate prediction of transport coefficients in the thermodynamic limit. It is observed in Fig.~\ref{Fig1} that the self-diffusion coefficient of LJ fluids is in excellent agreement with HS asymptote down to the lowest density studied in Ref.~\cite{Pieprzyk2019}. For the shear viscosity coefficient the HS asymptote somewhat overestimates the LJ data, but multiplying the HS asymptote by a constant factor of $0.9$ brings LJ and HS data to a very good agreement (see Fig.~\ref{Fig2}). For the thermal conductivity coefficients, the agreement between the LJ and HS data is good in the moderate density regime, $\rho_*/\rho_{*}^{\rm fr}\lesssim 0.6$, but for higher relative density the thermal conductivity coefficient of the HS fluid is consistently larger and overestimates that of the LJ fluid by about $40\%$ at the freezing point.        

It is not at all surprising that the HS model can reproduce the transport properties of softer particle systems. It is well recognized that properly reduced dynamical properties of fluids with strong repulsive forces can be mapped into those of HS fluids~\cite{Dymond1985,LopezFloresPRE2013,DyreJPCM2016}. The novelty of our approach is that the freezing density scaling seems to provide a simple and convenient mapping method, in particular for the self-diffusion coefficient.  

The fact that the freezing density scaling of LJ fluid is relatively close to that of a HS fluid is also not very surprising. The effective repulsive (inverse power law) exponent of a LJ potential is quite high, $n_{\mathrm{eff}}\simeq 18$ (at moderate densities)~\cite{PedersenPRL2008}. This is because the attractive ($\propto r^{-6}$) term of the LJ potential makes its repulsive short-range branch considerably steeper than just the $\propto r^{-12}$ repulsive term. Only at high densities and temperatures $n_{\mathrm{eff}}$ approaches 12~\cite{BohlingJCP2014}.
In this regime the agreement between the LJ and HS scaling may somewhat decline.

\begin{figure}
\includegraphics[width=8.5cm]{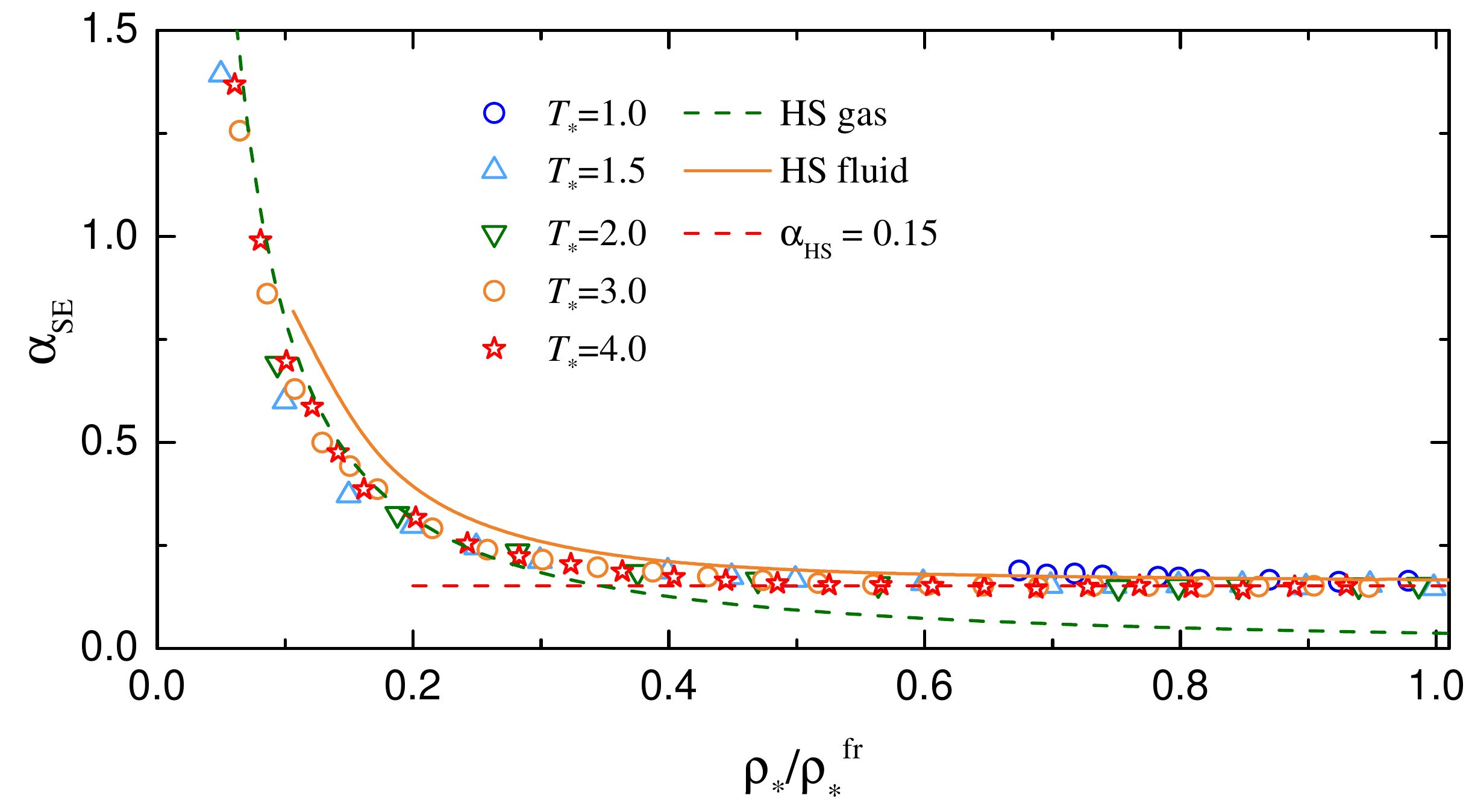}
\caption{(Color online) Stokes-Einstein relation without the hydrodynamic diameter in a LJ fluid. The parameter $\alpha_{\rm HS}$ is plotted as a function of the reduced density $\rho_*/\rho_*^{\rm fr}$. The symbols correspond to numerical data tabulated in Ref.~\cite{Meier2002,BaidakovFPE2011,BaidakovJCP2012}. The dashed and solid curves denote the HS gas and fluid asymptotes. The latter is based on the simulation data from Ref.~\cite{Pieprzyk2019}. The horizontal line at $\alpha_{\rm HS}= 0.15$, correspond to the simulation results from Ref.~\cite{CostigliolaJCP2019}.}
\label{Fig4}
\end{figure}    

Let us briefly discuss the Stokes-Einstein (SE) relation between the self-diffusion and viscosity coefficient. According to Zwanzig's model~\cite{ZwanzigJCP1983} for simple atomic fluids this relation can be formulated without introducing a so-called ``hydrodynamic diameter'' (in fact, the interparticle separation plays the role of the effective hydrodynamic diameter):
\begin{equation}\label{SE}
\frac{D\eta}{\rho^{1/3}T}=\alpha_{\rm SE}.
\end{equation} 
In the Rosenfeld's normalization Eq.~(\ref{SE}) reduces to $D_{\rm R}\eta_{\rm R}=\alpha_{\rm SE}$.
The parameter $\alpha_{\rm SE}$ can be related to the ratio of the transverse-to-longitudinal sound velocities~\cite{ZwanzigJCP1983,KhrapakMolPhys2019}:
\begin{displaymath}
\alpha_{\rm SE}\simeq 0.13\left(1+\frac{c_t^2}{2c_l^2}\right),
\end{displaymath}
where $c_t$ and $c_l$ are the transverse and longitudinal sound velocities. For LJ fluids near the freezing point we have $c_l\simeq 2 c_t$~\cite{KhrapakMolecules2020} and this should be relevant for a sufficiently dense regime not too far from the fluid-solid phase transition. Thus, we can expect $\alpha_{\rm SE}\simeq 0.15$ on approaching freezing. This is perfectly consistent with numerical data on the LJ fluids, see Fig.~\ref{Fig4}. The HS asymptote lies close, but slightly above the LJ data, because the HS viscosity is about $10\%$ higher than that of the LJ fluid. In the dilute regime we can expect an approach to the dilute HS gas asymptote, which is $D_{\rm R}\eta_{\rm R}\simeq 0.037/\rho_{*}^{4/3}$. Figure~\ref{Fig4} demonstrates that the data point lie quite close to this asymptote (as previously we used $\rho_*^{\rm fr}\simeq 1$ to plot this curve).           

The final comment is related to the crossover between the gas-like and liquid-like regions on the phase diagram, a topic of considerable recent interest~\cite{Simeoni2010,BrazhkinJPCB2011,BrazhkinPRE2012,
BrazhkinUFN2012,BellJCP2020}. Stokes-Einstein relation provides us with a good pragmatic tool to estimate the location of this crossover. Figure~\ref{Fig4} demonstrates that numerical data follow very closely the dilute gas and the fluid asymptotes. The later intersect at $\rho_*/\rho_*^{\rm fr}\simeq 0.35$ and this can be chosen as the demarcation condition between the gas-like and liquid-like regions. Note that this is close to the location of the minimum of the reduced shear viscosity coefficient (Fig.~\ref{Fig2}). It would be interesting to verify to which extent this condition is applicable to other soft interacting systems.        

\section{Conclusion}

To summarize, each of the properly reduced transport coefficients of self-diffusion, shear viscosity, and thermal conductivity of dense LJ fluids coincides along different isotherms when plotted as a function of density divided by its value at the freezing point. The quality of this coincidence is particularly high for the self-diffusion and viscosity coefficients and is somewhat lower for the thermal conductivity coefficient. The freezing density scaling is close to that of a HS fluid, in particular for the self-diffusion coefficient. This provides a simple useful procedure of mapping between soft (but strongly repulsive) and HS fluids. The Stokes-Einstein relation of the form $D_{\rm R}\eta_{\rm R}\simeq 0.15$ is satisfied to a very high accuracy in a dense fluid regime with $\rho_*/\rho_*^{\rm fr}\gtrsim 0.4$. At low densities the dilute HS asymptote is appropriate $D_{\rm R}\eta_{\rm R}\propto \rho_{*}^{-4/3}$. The two asymptotes are intersecting at about $\rho_*/\rho_*^{\rm fr}\simeq 0.35$ and this can serve as a practical demarcation condition between the gas-like and liquid-like regions on the LJ system phase diagram.        


\bibliography{TC_Ref}

\end{document}